\documentclass[twocolumn]{aastex63}
\usepackage{amsmath}
\usepackage{xcolor}

\begin{document}

\title{MAMMOTH-Subaru IV. Large Scale Structure and Clustering Analysis of Ly$\alpha$ Emitters and Ly$\alpha$ Blobs at $z=2.2-2.3$}

\author[0000-0003-2273-9415]{Haibin Zhang}
\affiliation{Department of Astronomy, Tsinghua University, Beijing 100084, People’s Republic of China; \url{haibin.zhang@nao.ac.jp, zcai@mail.tsinghua.edu.cn}}
\affiliation{National Astronomical Observatory of Japan, 2-21-1 Osawa, Mitaka, Tokyo 181-8588, Japan}

\author[0000-0001-8467-6478]{Zheng Cai}
\affiliation{Department of Astronomy, Tsinghua University, Beijing 100084, People’s Republic of China; \url{haibin.zhang@nao.ac.jp, zcai@mail.tsinghua.edu.cn}}

\author[0000-0001-6251-649X]{Mingyu Li}
\affiliation{Department of Astronomy, Tsinghua University, Beijing 100084, People’s Republic of China; \url{haibin.zhang@nao.ac.jp, zcai@mail.tsinghua.edu.cn}}

\author[0000-0002-2725-302X]{Yongming Liang}
\affiliation{National Astronomical Observatory of Japan, 2-21-1 Osawa, Mitaka, Tokyo 181-8588, Japan}

\author[0000-0003-3954-4219]{Nobunari Kashikawa}
\affiliation{National Astronomical Observatory of Japan, 2-21-1 Osawa, Mitaka, Tokyo 181-8588, Japan}
\affiliation{Department of Astronomy, School of Science, The University of Tokyo, 7-3-1 Hongo, Bunkyo-ku, Tokyo 113-0033, Japan}
\affiliation{Research Center for the Early Universe, The University of Tokyo, 7-3-1 Hongo, Bunkyo-ku, Tokyo 113-0033, Japan}

\author[0000-0002-0564-891X]{Ke Ma}
\affiliation{Department of Astronomy, Tsinghua University, Beijing 100084, People’s Republic of China; \url{haibin.zhang@nao.ac.jp, zcai@mail.tsinghua.edu.cn}}
\affiliation{School of Physics and Astronomy, University of Nottingham, University Park, Nottingham, NG7 2RD, UK}

\author[0000-0003-0111-8249]{Yunjing Wu}
\affiliation{Department of Astronomy, Tsinghua University, Beijing 100084, People’s Republic of China; \url{haibin.zhang@nao.ac.jp, zcai@mail.tsinghua.edu.cn}}
\affiliation{Steward Observatory, University of Arizona, 933 N Cherry Ave, Tucson, AZ 85721, USA}

\author[0000-0002-3119-9003]{Qiong Li}
\affiliation{Jodrell Bank Centre for Astrophysics, University of Manchester, Oxford Road, Manchester, UK}

\author[0000-0001-9487-8583]{Sean D. Johnson}
\affiliation{Department of Astronomy, University of Michigan, Ann Arbor, MI 48109, USA}

\author[0000-0003-3214-9128]{Satoshi Kikuta}
\affiliation{Department of Astronomy, School of Science, The University of Tokyo, 7-3-1 Hongo, Bunkyo-ku, Tokyo 113-0033, Japan}

\author[0000-0002-1049-6658]{Masami Ouchi}
\affiliation{Institute for Cosmic Ray Research, The University of Tokyo, Kashiwa, Chiba 277-8582, Japan}
\affiliation{National Astronomical Observatory of Japan, 2-21-1 Osawa, Mitaka, Tokyo 181-8588, Japan}
\affiliation{Kavli Institute for the Physics and Mathematics of the Universe (WPI), University of Tokyo, Kashiwa, Chiba 277-8583, Japan}

\author[0000-0003-3310-0131]{Xiaohui Fan}
\affiliation{Steward Observatory, University of Arizona, 933 N Cherry Ave, Tucson, AZ 85721, USA}

\author[0000-0001-9442-1217]{Yuanhang Ning}
\affiliation{Department of Scientific Research, Beijing Planetarium, Beijing 100044, China}

\begin{abstract}
We report the large scale structure and clustering analysis of Ly$\alpha$ emitters (LAEs) and Ly$\alpha$ blobs (LABs) at $z=2.2-2.3$. Using 3,341 LAEs, 117 LABs, and 58 bright (Ly$\alpha$ luminosity $L_{\rm Ly\alpha}>10^{43.4}$ erg s$^{-1}$) LABs at $z=2.2-2.3$ selected with Subaru/Hyper Suprime-Cam, we calculate the LAE overdensity to investigate the large scale structure at $z=2$. We show that 79\% LABs and 83\% bright LABs locate in overdense regions, which is consistent with the trend found by previous studies that LABs generally locate in overdense regions. We find that one of our 8 fields dubbed J1349 contains $39/117\approx33\%$ of our LABs and $22/58\approx38\%$ of our bright LABs. A unique and overdense $24'\times12'$ ($40\times20$ comoving Mpc$^2$) region (J1347 protocluster) has 12 LABs (8 bright LABs). By comparing to SSA22 that is one of the most overdense LAB regions found by previous studies, we show that the J1347 protocluster region has a higher bright LAB density than the SSA22 protocluster region with a $1\sigma$ significance. We calculate the angular correlation functions (ACFs) of LAEs and LABs in the unique J1349 field and fit the ACFs with a power-law function to measure the slopes. The bright LABs show a $5\sigma$ larger slope suggesting that bright LABs are more clustered than faint LAEs. Our LABs have a large galaxy bias of $\sim 5-7$, which suggests that LABs generally reside in more massive dark matter halos (halo masses $M \gtrsim 10^{13}$ M$_{\odot}$) than faint LAEs. 

\end{abstract}


\section{Introduction} \label{sec:intro}
In the last two decades, luminous and spatially extended 
Ly$\alpha$ emitters (LAEs) have been identified at $z=2-7$ by 
wide-field imaging surveys. These luminous and extended LAEs are 
often referred to as Ly$\alpha$ blobs (LABs; e.g. \citealt{mccarthy1987, keel1999, steidel2000, Haiman2000, francis2001, Reuland2003, matsuda2004, dey2005, Nilsson2006, ouchi2009, yang2009, Hennawi2009, Prescott2009, Faucher2010, Zafar2011, Prescott2012, Prescott2013, Bridge2013, Cantalupo2014, sobral2015, Hennawi2015, cai2017, Battaia2018, shibuya2018, kikuta2019, zhang2020, Hibon2020, Jim2023, li2024}). 
The Ly$\alpha$ luminosities of LABs are commonly brighter than $\sim10^{43}$ erg s$^{-1}$, and the sizes are much larger than point sources. The luminous and extended nature of LABs allows the detection of Ly$\alpha$ emission in the circumgalactic medium (CGM) scale at high-$z$ even with a limited imaging depth. Because Ly$\alpha$ emission traces the hydrogen gas in the CGM (\citealt{Byrohl2023}), it is widely believed that LABs are important objects to study formation and evolution of galaxies in the early Universe. 

Previous studies have shown that LABs commonly locate in overdense regions, 
suggesting that the formation of LABs is affected by the environment such as 
the large scale structure (e.g. \citealt{francis1996,steidel2000,yang2009, Kollmeier2010,matsuda2011,kikuta2019,zhang2020,Ramakrishnan2023,Ramakrishnan2024, Lee2024}).
The field-to-field variation of LABs is found to be strong (\citealt{yang2010}). 
One of the densest LAB fields found by previous studies is the SSA22 field at $z=3.1$. 
In a $31'\times23'$ ($\sim 58\times43$ cMpc$^2$) region of the SSA22 field, 35 
LABs have been identified (\citealt{matsuda2004}). Because such a dense LAB field 
is very rare, the discovery strongly depends on wide-field surveys and the 
identification of new dense LAB fields provides important insight on the 
extreme galaxy formation environment. 

Clustering analysis using methods such as the angular correlation function (ACF) is important for understanding the spatial distribution of galaxies. Although the ACF of LAEs has been investigated at high-$z$ (e.g. \citealt{ouchi2010,harikane2016}), 
limited information about the 
ACF of LABs 
can be found from the literature yet. By comparing the ACFs of LAEs and LABs, 
one can understand if the underlying dark matter halos affect the formation of 
LAEs and LABs similarly. 

In this paper, we investigate the large scale structure and carry out clustering 
analysis of LAEs and LABs at $z=2.2-2.3$. This paper is organized as follows. 
In Section \ref{sec:data}, we introduce our observations and data reduction. 
Using the reduced data, we show our selection of LAEs and LABs in Section \ref{sec:selection}. The results and discussion are presented in Section \ref{sec:results}. Finally, we summarize this paper in Section \ref{sec:summary}.
The AB magnitudes \citep{oke1983} and physical distances are used throughout this paper unless indicated otherwise. We adopt the $\Lambda$CDM cosmology with $\Omega_m=0.3$, $\Omega_\Lambda=0.7$, and $h=0.7$.

\section{Observations and Data Reduction} \label{sec:data}

The imaging data we use in this study are obtained with Subaru/Hyper Suprime-Cam 
(HSC; \citealt{miyazaki2018, komiyama2018, kawanomoto2018, furusawa2018}). 
We carry out observations using narrowband (NB387 and NB400) and broadband ($g$) 
filters between January 2018 and March 2020. The NB387, NB400, and $g$ filters 
are centered at 3863, 4003, and 4754 {\AA}, respectively. The filter widths (FWHM) 
of NB387, NB400, and $g$ are 55, 92, and 1395 {\AA}, respectively. 
The observations are carried out in 8 fields with a total survey area of $\sim12$ deg$^2$. 
Details of the field selection are presented in \citet{cai2016}, 
\citet{liang2021}, Cai et al. (in prep.), and Liang et al. (in prep.). 
The seeing sizes are between $0''.7$ and $1''.2$. We summarize the information 
of the 8 fields in Table \ref{tab:fields}.

We reduce the imaging data with the HSC pipeline dubbed \textit{hscPipe} (\citealt{bosch2018, aihara2018}). 
Details of the data reduction in NB387 and NB400 fields are presented in \citet{liang2021} and \citet{zhang2024}, respectively. 
After data reduction, we measure the detection limits in a $1''.7$ diameter aperture, 
except for J0210 field in a $2''.5$ diameter aperture because the seeing in J0210 
is the worst. The $5\sigma$ detection limits are $24.3-25.0$, $25.5-25.8$, 
and $26.2-27.0$ mag for NB387, NB400, and $g$, respectively. 
The source detection and photometry are carried out with SExtractor \citep{bertin1996} 
and described in \citet{liang2021} and \citet{zhang2024}, after matching the point-spread functions (PSFs) of the NB and BB filters with proper Gaussian kernels. 
We do not use regions such as those near bright stars and field edges because of the 
low signal-to-noise ratios (SNRs).

We visually inspect the reduced image with S/N (signal to noise ratio) maps and bright star catalogs (e.g. GAIA), and manually mask out saturated and low S/N regions with circles and boxes around bright stars, in the same manner as \citet{liang2021}.
The sizes of these masked regions depend on the brightness of stars, and the radius of circular masks is typically a few ten arcsec. But the radius of masks can be as large as a few hundred arcsec if the star is very bright, and these large masks can be easily seen in Figure \ref{fig:delta} as the white regions.

\begin{deluxetable*}{ccccccccccccc}
\tablenum{1}
\tablecaption{Information of 8 fields in this study\label{tab:fields}}
\tablewidth{0pt}
\tablehead{
\colhead{Field} & \colhead{Filters} & \colhead{$\theta_{\mathrm{NB}}$} & \colhead{$\theta_{\mathrm{BB}}$} & \colhead{$m_{\mathrm{NB}, 5 \sigma}$} & \colhead{$m_{\mathrm{BB}, 5 \sigma}$} & \colhead{Ly$\alpha$ limit} & \colhead{IsoArea$_{\rm min}$} & \colhead{Area} &\colhead{$N_{\rm LAE}$} & \colhead{$N_{\rm LAB}$} & \colhead{$N_{\rm LAB, brt}$} & \colhead{$\Sigma_{\rm LAB, brt}$} \\
\colhead{name} & \colhead{name} & \colhead{arcsec} & \colhead{arcsec} &
\colhead{mag} & \colhead{mag} & \colhead{see note} & \colhead{arcsec$^2$} & \colhead{deg$^2$} & count &count &count & degree$^{-2}$
}
\decimalcolnumbers
\startdata
J0210 & NB387 \& $g$ & 1.22 & 0.90 & 24.25 & 26.34 & 10.56 &12 & 1.34  & 227 & 10 & 5 &3.73 \\
J0222 & NB387 \& $g$ & 0.90 & 0.90 & 24.99 & 27.01 & 8.23 &16 & 1.13  & 422 & 10 & 4 &3.54 \\
J0924 & NB387 \& $g$ & 0.84 & 0.79  & 24.74 & 26.63 & 9.79 &12 & 1.47  & 311 & 5  & 2 &1.36 \\
J1419 & NB387 \& $g$ & 0.86 & 0.70 & 24.81 & 26.80 & 10.66 &12 & 1.45 & 280 & 12 & 3 &2.07  \\
J0240 & NB400 \& $g$ & 1.00 & 0.84  & 25.61 & 26.80 & 7.78 &16 & 1.53  & 517 & 24 & 16 &10.46 \\
J0755 & NB400 \& $g$ & 0.88 & 1.16  & 25.83 & 26.50 & 5.04 &25 & 1.54 & 545 & 9 & 4 &2.60 \\
J1133 & NB400 \& $g$ & 0.85 & 0.82  & 25.49 & 26.30 & 9.19 &16 & 1.55 & 403 & 8 & 2 &1.29 \\
J1349 & NB400 \& $g$ & 0.98 & 0.86  & 25.67 & 26.15 & 7.76 &16 & 1.62 & 636 & 39 & 22 &13.58 \\
\enddata
\tablecomments{ Column 1: field name; Column 2: filter name; Column 3: seeing of narrowband in FWHM; Column 4: seeing of broadband in FWHM; Column 5: 5$\sigma$ limiting magnitude of narrowband; Column 6: 5$\sigma$ limiting magnitude of broadband; Column 7: average 2$\sigma$ detection limit of assumed Ly$\alpha$ emission in $10^{-18}$ erg s$^{-1}$ cm$^{-2}$ arcsec$^{-2}$; Column 8: minimum isophotal area for our LAB selection; Column 9: effective survey area after masking; Column 10: number of LAEs selected in this study: Column 11: number of LABs selected in this study; Column 12: number of bright LABs selected with Ly$\alpha$ luminosities $L_{\rm Ly\alpha} > 10^{43.4}$ erg s$^{-1}$} in this study; Column 13: surface number density of bright LABs selected in this study. 
\end{deluxetable*}

\section{Sample Selection} \label{sec:selection}
Similar to previous studies (e.g. \citealt{konno2016, liang2021}), we select LAEs with a narrowband minus broadband (NB-BB) color excess that corresponds to a rest-frame Ly$\alpha$ equivalent width of $>20${\AA}. 
Details of our LAE selection are presented in \citet{liang2021}, \citet{zhang2024}, and \citet{ma2024}. 
In brief, NB387 LAEs are selected with the following criteria:
\begin{equation} \label{eq:NB387}
\begin{gathered}
g_{\rm ap}-NB387_{\rm ap}>0.3\ \mathrm{and}\ g_{\rm tot}-NB387_{\rm tot}>0.3\ \mathrm{and}\\
g_{\rm ap}-NB387_{\rm ap}>(g_{\rm ap}-NB387_{\rm ap})^{3\sigma}\ \mathrm{and}\\
g_{\rm tot}-NB387_{\rm tot}>(g_{\rm tot}-NB387_{\rm tot})^{3\sigma}\ \mathrm{and}\\
20.5<NB387_{\rm ap}<NB387^{5\sigma},
\end{gathered}
\end{equation}
where the subscripts ``ap" and ``tot" are the aperture and total magnitudes, respectively. 
The ``AUTO'' magnitude in SExtrator is used as the total magnitude. 
For sources whose $g$ magnitudes are fainter than the $2\sigma$ limit, 
we use the $2\sigma$ limit instead. We calculate the $3\sigma$ color uncertainty 
$(g-NB387)^{3\sigma}$ by $(g-NB387)^{3\sigma}=-2.5\log(1\pm3\sqrt{f_{\rm{err},g}^2+f_{\rm{err},\rm{NB387}}^2}/f_{\rm NB387})$, 
where $f_{\rm{err},g}$ and $f_{\rm{err},\rm{NB387}}$ are the $1\sigma$ uncertainties 
in $g$ and NB387, respectively. We then visually inspect sources 
passing the criteria and remove spurious sources such as cosmic rays 
and satellite trails. Our final NB387 LAE catalog contains 1,240 candidates.

Similarly, NB400 LAEs are selected with the following criteria:
\begin{equation} \label{eq:NB400}
\begin{gathered}
g_{\rm ap}-NB400_{\rm ap}>0.4\ \mathrm{and}\ g_{\rm tot}-NB400_{\rm tot}>0.4\ \mathrm{and}\\
g_{\rm ap}-NB400_{\rm ap}>(g_{\rm ap}-NB400_{\rm ap})^{3\sigma}\ \mathrm{and}\\
g_{\rm tot}-NB400_{\rm tot}>(g_{\rm tot}-NB400_{\rm tot})^{3\sigma}\ \mathrm{and}\\
18.0<NB400_{\rm ap}<NB400^{5\sigma},
\end{gathered}
\end{equation}
where we use the same symbols as Equation \ref{eq:NB387}. 
We carry out visual inspection after applying the selection criteria to remove spurious sources such as cosmic rays and satellite trails. 
Our final NB400 LAE catalog contains 2101 candidates after visual inspection. 
The number of LAE candidates in each field is presented in Table \ref{tab:fields}.

\begin{figure}[htb]
\plotone{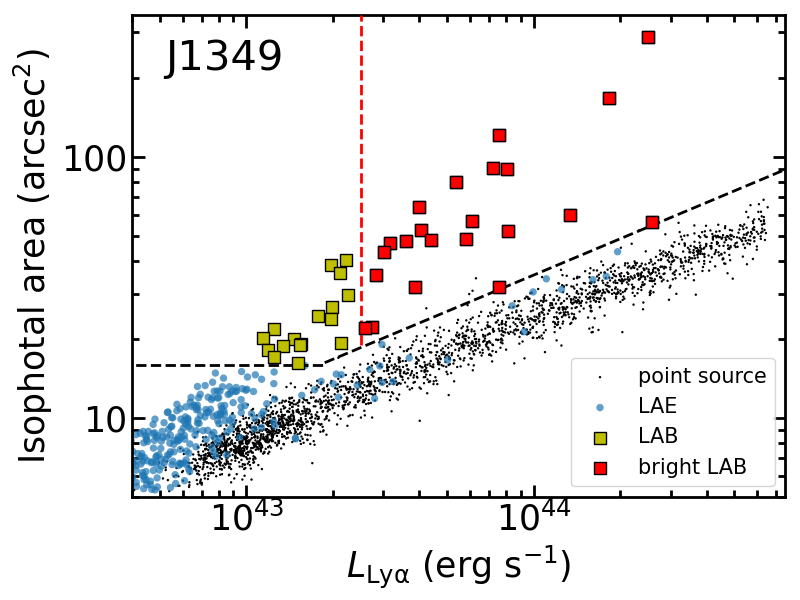}
\caption{Isophotal area as a function of Ly$\alpha$ luminosity $L_{\rm Ly\alpha}$ of 
our LAEs (blue filled circle), LABs (yellow filled square), and bright LABs 
(red filled square). Black dots are point sources. For these point sources, we assume that all of the flux in the NB comes from Ly$\alpha$ emission. Black dashed lines show the 
criteria we use to select LABs. 
Among the LABs, we select bright LABs with $L_{\rm Ly\alpha}>10^{43.4}$ erg s$^{-1}$ 
(red dashed line). This figure shows our selection in the J1349 field as an example 
of the 8 fields. Our selection in the other 7 fields are presented in \citet{li2024}. }
\label{fig:LAB_selection}
\end{figure}

After obtaining the LAE catalogs, we select LABs based on the Ly$\alpha$ 
luminosity $L_{\rm Ly\alpha}$ and isophotal area as presented in detail in Li et al. 
(\citeyear{li2024}; Figure \ref{fig:LAB_selection}). 
We calculate Ly$\alpha$ luminosities with NB and BB magnitudes assuming a flat  continuum emission (with a constant magnitude) and constant redshifts correspond to the NB centers. If the Ly$\alpha$ line of a source offsets from the NB center, the Ly$\alpha$ flux and luminosity may be underestimated. Detailed methods and formulae of the calculation are presented in the Appendix of \citet{li2024}.
We define the isophotal area 
using the area above the 2$\sigma$ surface brightness limit. The average 2$\sigma$ surface brightness limit is shown in Table \ref{tab:fields}, and is estimated using the pixel-to-pixel variation in sky regions (regions with no detected sources) of the Ly$\alpha$ (continuum-subtracted) image. We do not use circular apertures for this estimation because our isophotal area measurement is based on pixels but not apertures. We select 
LABs with an isophotal area larger than 3$\sigma$ confidence levels 
of the point sources. For these point sources, we assume that all of the flux in the NB comes from Ly$\alpha$ emission, to estimate the point spread function (PSF). The isophotal area should also be larger than a minimum 
value that depends on the depth of each field, assuming a surface brightness profile of the model in \citet{kimock2021}. The minimum isophotal areas are shown in Table \ref{tab:fields}. 

After applying the criteria, we carry out visual inspection to remove sources whose isophotal areas are affected by adjacent 
artifacts or bright stars. After visual inspection, we select 37 and 80 LAB 
candidates at $z=2.2$ and 2.3, respectively. Among these LAB candidates, 
we further select LABs with Ly$\alpha$ luminosities $L_{\rm Ly\alpha}$ 
larger than $10^{43.4}$ erg s$^{-1}$, and these LABs are referred to as bright LABs hereafter. The selection of bright LABs is an empirical criterion adopted by previous studies (e.g. \citealt{shibuya2018,zhang2020}), and we use this criterion for a comparison in Section \ref{sec:results}. The numbers of bright LABs are 14 and 44 at $z=2.2$ and 2.3, respectively. 
The number of LABs in each field is presented in Table \ref{tab:fields}. 

Because the effective areas and detection limits are not the same for each field, in later sections we only compare the number densities of bright LABs which are selected using the same method (same Ly$\alpha$ luminosity threshold and 3$\sigma$ extension) across our 8 fields. 

We have carried out spectroscopic follow-up observations of more than 100 LAE candidates with Magellan/IMACS. The contamination rate estimated from the spectra is $\lesssim8\%$, as presented in detail in \citet{zhang2024}. We have removed all spectroscopically confirmed low-z interlopers from our sample with our observations and archival data. The remaining bright compact sources (e.g. bright blue dots in Figure 1) are generally QSOs. The {\sc [O ii]} emitters at $z<0.1$ are expected to be the largest contaminants for our $z=2$ LAEs. Because not all of our LAE candidates have been spectroscopically observed, we cannot exclude all {\sc [O ii]} emitters. However, the contamination rate contributed by these {\sc [O ii]} emitters in the literature is low (e.g., $<3\%$ in \citealt{cooper2023}), and we expect that their contribution to our results is negligible.

The detection completeness of LAEs is estimated by the Monte Carlo method, as presented in our series paper \citet{ma2024}. In brief, we generate mock galaxies with NB magnitudes of $20–27$ mag using the GALSIM package (\citealt{rowe2015}). Source detection and photometry of the mocks are carried out in the same manner as our LAEs. We find that the completeness is typically above 90\% for bright LAEs (NB387$>23.5$ mag or NB400$>24.0$ mag), and falls rapidly to 0 beyond the 5$\sigma$ depths. On the other hand, it should be noted that our selection of LABs may not be complete especially at the faint end, because a source may not be selected as a LAB if its radial profile is more compact than our assumed model of \citet{kimock2021}.

\section{Results and Discussion} \label{sec:results}

\subsection{LAE Overdensity}

\begin{figure*}[htb]
\includegraphics[width=0.95\textwidth]{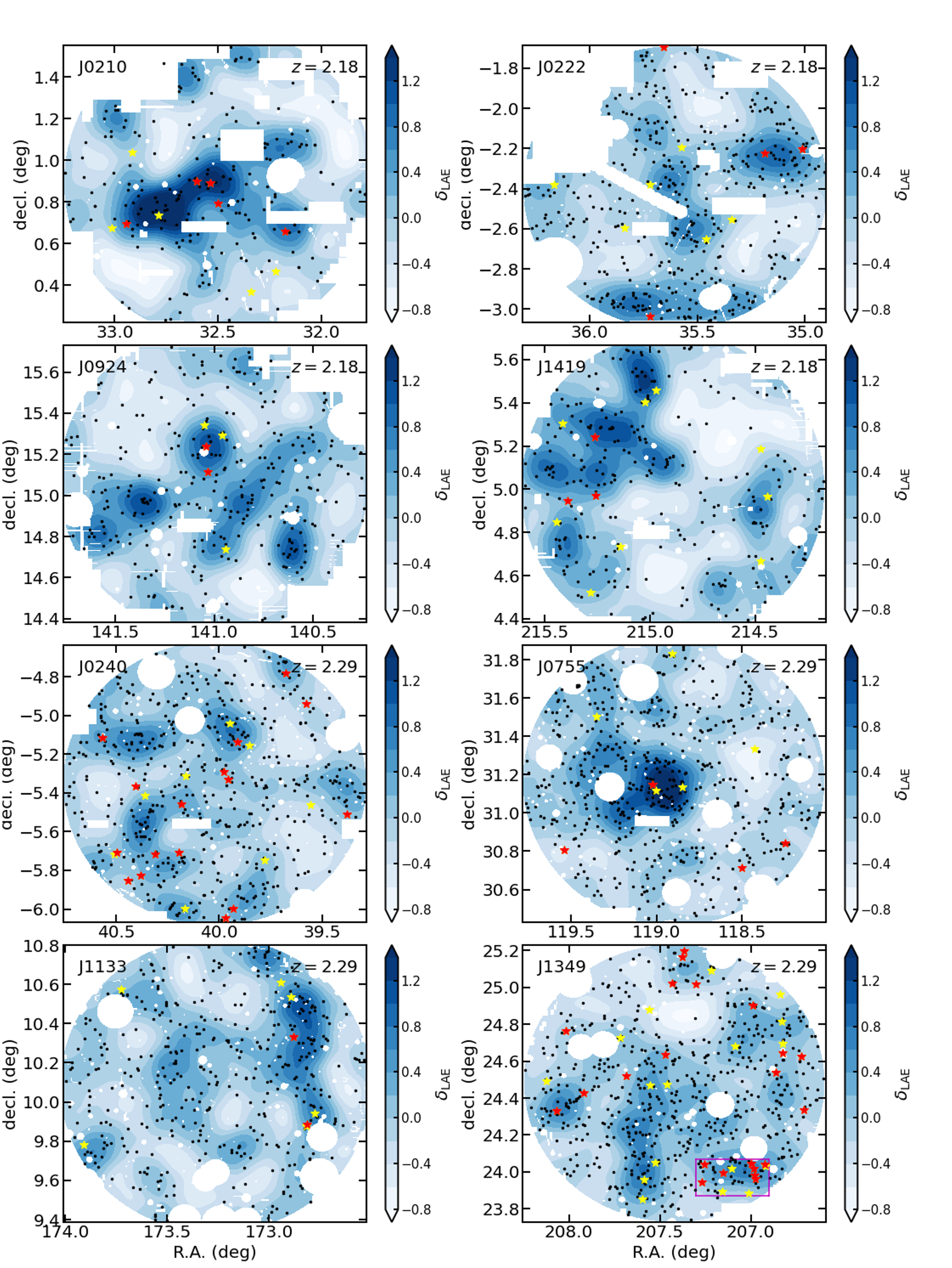}
\caption{LAE overdensity maps of the 8 fields in this study. The black dots, yellow stars, and red stars are our LAEs, LABs, and bright LABs, respectively. The LAE overdensity is represented by the blue colored shade, and a darker blue color means a higher overdensity. Masked regions are shown in white. The magenta box indicates the J1347 protocluster region. Because different $\bar{N}$ values are used in different fields, the overdensity in different fields should not be directly compared.}
\label{fig:delta}
\end{figure*}

To investigate the large scale structure in our 8 fields, we calculate the 
LAE overdensity. The LAE overdensity $\delta_{\rm LAE}$ is defined by
\begin{equation} \label{eq:delta}
\begin{gathered}
\delta_{\rm LAE}=\frac{N-\bar{N}}{\bar{N}}, \\
\end{gathered}
\end{equation}
where $N$ and $\bar{N}$ are the number and field-averaged number of LAEs 
in a circular aperture, respectively. The $\bar{N}$ is calculated in each field separately. The diameter of the aperture is 20 comoving 
Mpc (cMpc) that is consistent with previous studies at $z=2$ (e.g. \citealt{liang2021}). 
We assume the Poisson statistics for $N$ and $\bar{N}$ in our calculation. Before counting the number of LAEs with the circular aperture, we fill the masked regions (white regions in Figure \ref{fig:delta}) with the mean LAE number densities in each field.
Figure \ref{fig:delta} shows the LAE overdensity map of the 8 fields 
along with our LABs. We find that 92 out of 117 LABs and 48 out of 58 bright LABs locate in overdense regions with $\delta_{\rm LAE}>0$. The overdense fractions are 92/117=79\% and 48/58=83\% for LABs and bright LABs, respectively. This result is consistent with the trend found by previous studies that LABs generally locate in overdense regions (e.g. \citealt{badescu2017,kikuta2019,zhang2020}). Additionally, we fit a Gaussian function to the $\delta_{\rm LAE}$ distribution in each field and measure the $\sigma$. We count the LAB numbers in regions with $\delta_{\rm LAE}>2\sigma$, and find that 15\% of LABs and 17\% of bright LABs locate in these high overdensity regions.

\begin{figure*}[htb]
\includegraphics[width=\textwidth]{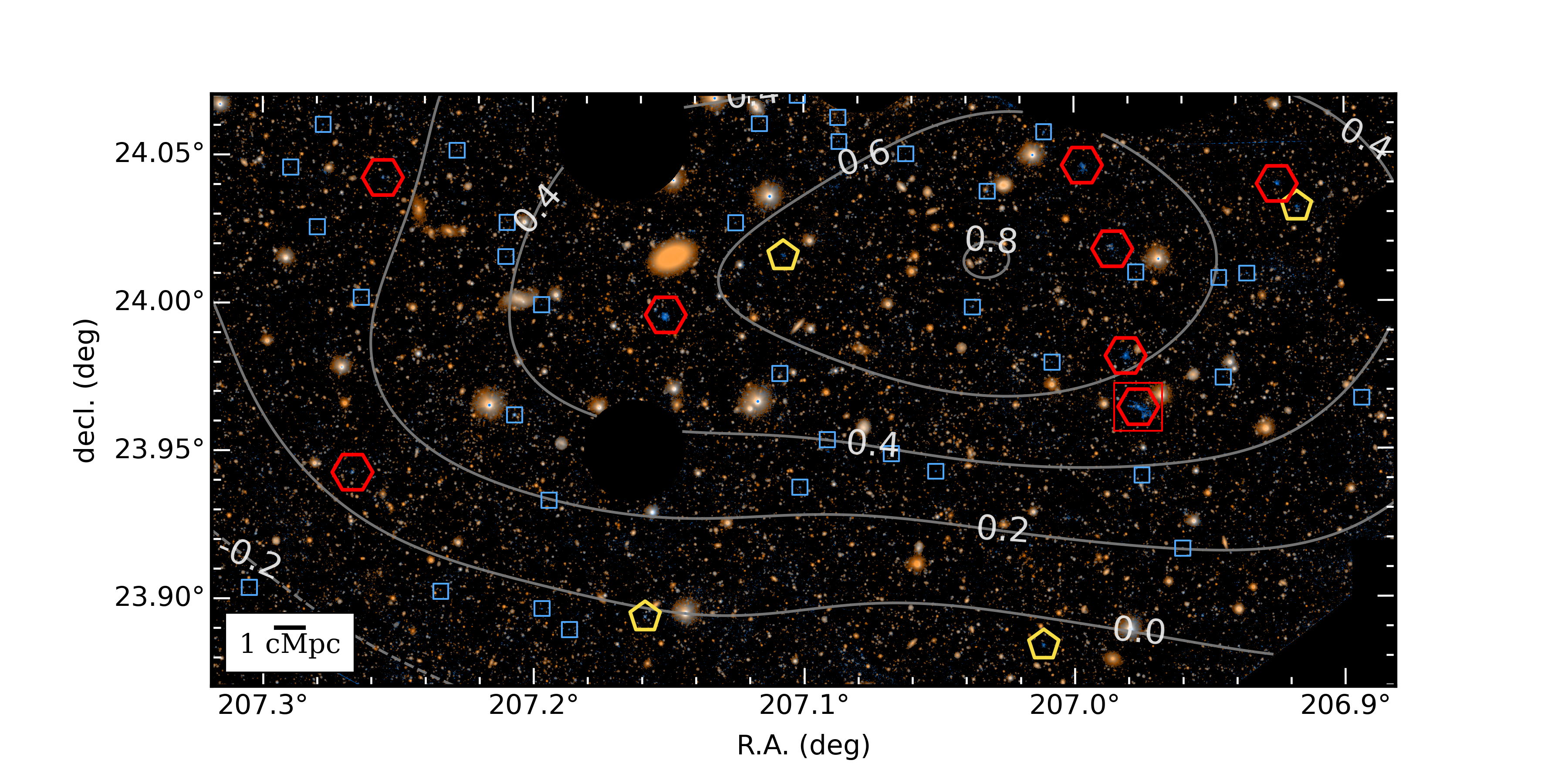}
\caption{Pseudo color image of the J1347 protocluster region. The blue and red color layers are presented using NB400 and $g$ images, respectively. LAEs, LABs, and bright LABs are indicated by blue, yellow, and red boxes, respectively. The gray contours show the overdensity levels indicated by white numbers. }
\label{fig:J1349_pseudo}
\end{figure*}

Interestingly, we find that the J1349 field contains $39/117\approx33\%$ of our 
LABs and $22/58\approx 38\%$ of our bright LABs. The bright LAB number density in J1349 is also the highest and cannot be explained by the Poisson statistics (Table 1).
The J1349 field shows an overdense $24'\times12'$ ($\sim 40\times20$ cMpc$^2$) 
region (Figure \ref{fig:J1349_pseudo}) that has 12 LABs (8 bright LABs). This $24'\times12'$ region is referred to as the J1347 protocluster region hereafter. 

We compare the J1347 protocluster region with one of the most overdense LAB regions found by previous studies. Matsuda et al. (\citeyear{matsuda2004}; hereafter M04) identify 35 LABs in 
a $31'\times23'$ ($\sim 58\times43$ cMpc$^2$) region of the SSA22 field at $z=3.1$. 
Four out of these 35 LABs are bright LABs with $L_{\rm Ly\alpha}>10^{43.4}$ erg s$^{-1}$. Because M04 use different selection criteria (isophotal area at a different detection limit; $>16$ arcsec$^2$ above $2.2 \times 10^{-18}$ erg s$^{-1}$ cm$^{-2}$ arcsec$^{-2}$) at $z=3.1$ from our selection, 
it is not fair to directly compare the LAB numbers. For fair comparison, we convert the criteria in M04 to our observations. The detection limit would be $5.2 \times 10^{-18}$ erg s$^{-1}$ at $z=2.3$ assuming the surface brightness dimming of $(1+z)^{-4}$. It should be noted that this limit is deeper than our limit in J1349 ($7.76 \times 10^{-18}$ erg s$^{-1}$), and that our LABs would appear larger if observed to the same depth as the M04 observations. Because the LABs in J1349 are also selected with a minimum area of $16$ arcsec$^2$, our LABs would also satisfy the selection criteria in M04.
Finally, we compare the number densities of bright LABs in J1347 and SSA22 protocluster regions. 
The survey volumes of the J1347 and SSA22 are $7.3\times10^4$ and $1.3\times10^5$ cMpc, respectively.  The volume densities of bright LABs are then $1.1\times10^{-4}$ and $3.1\times10^{-5}$ cMpc$^{-3}$ for the J1347 and SSA22 protocluster regions, respectively. If we use a same survey volume of $1.3\times10^5$ cMpc, the volume density of bright LABs in the J1347 protocluster region becomes $6.2\times10^{-5}$ cMpc$^{-3}$.
Assuming the Poisson statistics in \citet{gehrels1986}, the J1347 protocluster region has a higher bright LAB density than the SSA22 protocluster region with a $1\sigma$ significance.

\subsection{Angular Correlation Function}

To further investigate the unique J1349 field, we calculate the angular correlation functions (ACFs) 
of LAEs and LABs. We calculate the ACF $\omega(\theta)$ with the following equation:
\begin{equation} \label{eq:omega}
\begin{gathered}
\omega(\theta)=\frac{DD(\theta)-2DR(\theta)+RR(\theta)}{RR(\theta)}, \\
\end{gathered}
\end{equation}
where $DD(\theta)$, $DR(\theta)$, and $RR(\theta)$ are the numbers of galaxy-galaxy, 
galaxy-random, and random-random pairs divided by the total number of pairs in galaxy-galaxy, 
galaxy-random, and random-random samples, respectively (\citealt{landy1993}). 
The random sample contains 600,000 randomly generated points, which correspond to a surface density of $\sim 100$ arcmin$^{-2}$. This surface density is consistent with the random catalogs (\citealt{aihara2019}) used by previous clustering studies with Subaru/HSC (e.g. \citealt{ouchi2018}). The masked regions are taken into account when generating the random points, while the LAB and LAE selection criteria are not included as the random points do not have photometric information. We use the same random sample to calculate ACFs. 
We assume that $DD(\theta)$, $DR(\theta)$, and $RR(\theta)$ pair counts follow the Poisson statistics, and calculate the propagated error to $\omega(\theta)$ with common error propagation formulae. Because the numbers of LABs are small, we use the small number statistics presented in \citet{gehrels1986}.

\begin{figure*}[htb]
\plotone{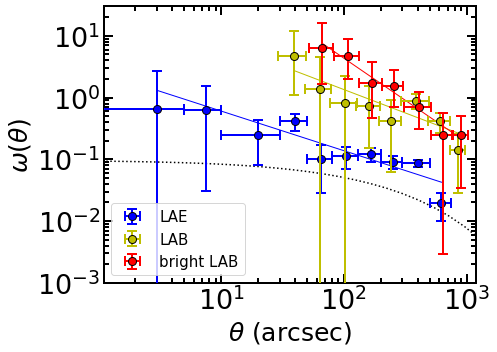}
\caption{Angular correlation functions of LAEs (blue), LABs (yellow), and bright LABs (red) in the J1349 field. The solid lines show the best-fit power-law functions of the ACFs, while the dotted line presents the ACF of dark matter at $z=2.3$. The best-fit slopes of the ACFs are $0.64\pm 0.14$ , $0.76\pm 0.26$, and $1.36\pm 0.11$ for LAEs, LABs, and bright LABs, respectively. The horizontal ranges of the solid lines are equal to the ranges of data points we used for the fitting. To avoid overlaps, the red and yellow points and lines are shifted along the horizontal axis by +0.01 and -0.01 dex, respectively.}
\label{fig:ACF}
\end{figure*}

For the calculation of ACFs, we exclude all LABs from the LAE sample, while the LAB sample include bright LABs given the small LAB numbers. Figure \ref{fig:ACF} shows the ACFs of our LAEs and LABs in the J1349 field. 
We fit the ACFs with a power-law function $\omega(\theta)=A_{\omega} \theta^{-\beta}$ and measure the amplitude $A_{\omega}$ and slope $\beta$. The best-fit parameters are summarized in Table \ref{tab:ACF}. The errors of parameters are the uncertainties from the least squares fitting. We find that the slopes of LAEs and LABs are similar within $1\sigma$, while the bright LABs show a $5\sigma$ larger slope than LAEs.
This result suggests that bright LABs are more clustered than LAEs, consistent with our K-function measurements presented in the Appendix.
On the other hand, \citet{herrero2023} find that luminous LAEs are more clustered and reside in more massive dark matter halos than faint LAEs. We cannot conclude if bright LABs are a different population from LAEs based on the strong clustering.

\begin{deluxetable}{cccc}
\tablenum{2}
\tablecaption{Best-fit parameters of ACFs \label{tab:ACF}}
\tablewidth{0pt}
\tablehead{
\colhead{Sample} & \colhead{$A_{\omega}$} & \colhead{$\beta$} & \colhead{$b_{g}$} 
}
\decimalcolnumbers
\startdata
LAE & 2.65 $\pm$ 2.02 & 0.64 $\pm$ 0.14 & 1.77 $\pm$ 0.09 \\
LAB & 44.79 $\pm$ 70.62 & 0.76 $\pm$ 0.26 & 5.31 $\pm$ 0.53  \\
Bright LAB & $(2.24 \pm 1.39)\times10^{3} $ & 1.36 $\pm$ 0.11 & 6.02 $\pm$ 0.98 \\
\enddata
\tablecomments{Column 1: sample name; Column 2: amplitude of power-law function; Column 3: slope of power-law function; Column 4: galaxy bias. }
\end{deluxetable}

We estimate the galaxy bias $b_{g}$ with the following method. First, we calculate the spatial correlation function of dark matter $\xi(r,z)$ using the ``Colossus" package (\citealt{diemer2018}) with the dark matter power spectrum model in \citet{eisenstein1998}, where $r$ is the radius and $z$ is the redshift. We assume the $\Lambda$CDM cosmology with $\Omega_m=0.3$, $\Omega_\Lambda=0.7$, and $h=0.7$. We then calculate the dark matter ACF $\omega_{DM}(\theta)$ with the Limber's equation (\citealt{limber1953}). Finally, the galaxy bias is calculated with $b_{g} = \sqrt{\omega_{g}(\theta)/\omega_{DM}(\theta)}$, where $\omega_{g}(\theta)$ and $\omega_{DM}(\theta)$ are the ACFs of galaxies and dark matter, respectively. We use the inverse square of errors in $b_{g}$ (propagated from $\omega_{g}(\theta)$) as weights, and calculate the weighted average $b_{g}$. The $b_{g}$ of LAEs, LABs, and bright LABs are presented in Table \ref{tab:ACF}. 

While the galaxy bias of LAEs is consistent with previous studies (e.g. \citealt{ouchi2020, White2024}), we find that LABs (and bright LABs) show a large galaxy bias of $\sim 5-7$. Because the cosmic variance $\sigma_{\rm g}$ 
is given by $\sigma_{\rm g}=b_g \sigma_{\rm DM}$, 
where $\sigma_{\rm DM}$ is the density fluctuation of dark matter and 
is expected to be nearly constant for the same redshift and survey volume, 
the large bias causes a large $\sigma_{\rm g}$ of LABs. 
The large $\sigma_{\rm g}$ is consistent with the strong field-to-field 
variance of bright LAB number densities found in this study (Table 1) that cannot be explained by the Poisson statistics. Interestingly, previous studies also find that LABs have a strong field-to-field variance, and based on N-body simulations suggest a bias of $\sim 7$ (\citealt{yang2010}) that is consistent with our measurement of bright LABs.

The large galaxy bias of LABs also suggests that LABs generally reside in more massive dark matter (DM) halos than faint LAEs. According to the relations between galaxy bias, stellar mass, and halo mass (e.g. \citealt{moster2010}), our LABs generally reside in DM halos with halo masses $M \gtrsim 10^{13}$ M$_{\odot}$, which are heavier than those of our LAEs typically ($M \sim 10^{11}$ M$_{\odot}$). Consistently, \citet{yang2010} also suggest halo masses of $M \gtrsim 10^{13}$ M$_{\odot}$ for LABs at $z=2.3$ based on N-body simulations, and these halos likely evolve to $M \sim 10^{14}$ M$_{\odot}$ at present-day. Additionally, previous studies estimate the total masses derived from galaxy overdensities in protoclusters, and show that protoclusters with LABs tend to have heavier total masses (mostly $\sim 10^{14}-10^{15}$ M$_{\odot}$ at present-day) than protoclusters without LABs (\citealt{Ramakrishnan2023}). Our results may also suggest a trend that bright LABs (with $L_{\rm Ly\alpha}>10^{43.4}$ erg s$^{-1}$) likely reside in more massive dark matter halos than the general LAB population.

\section{Summary} \label{sec:summary}
In this study, we investigate the large scale structure and carry out clustering analysis of LAEs and LABs at $z=2.2-2.3$. Our results are summarized below.

\begin{enumerate}
    \item Using 3341 LAEs, 117 LABs, and 58 bright (Ly$\alpha$ luminosity $L_{\rm Ly\alpha}>10^{43.4}$ erg s$^{-1}$) LABs at $z=2.2-2.3$, we calculate the LAE overdensity to investigate the large scale structure at $z=2$. We show that 79\% LABs and 83\% bright LABs locate in overdense regions with $\delta_{\rm LAE}>0$, which is consistent with the trend found by previous studies that LABs generally locate in overdense regions. Additionally, 15\% of LABs and 17\% of bright LABs locate in high overdensity regions with $\delta_{\rm LAE}>2\sigma$.

    \item We find that the J1349 field contains $39/117\sim33\%$ of our LABs and $22/58\sim38\%$ of our bright LABs. A unique and overdense $24'\times12'$ ($\sim 40\times20$ comoving Mpc$^2$) region (J1347 protocluster region) in J1349 has 12 LABs (8 bright LABs). By comparing to SSA22 that is one of the most overdense LAB regions found by previous studies, we show that the J1347 protocluster region has a higher bright LAB density than the SSA22 protocluster region with a $1\sigma$ significance.

    \item We calculate the angular correlation functions of LAEs and LABs in the unique J1349 field and fit the ACFs with a power-law function to measure the slopes. The best-fit slopes of LAEs, LABs, and bright LABs are $0.64\pm 0.14$ , $0.76\pm 0.26$, and $1.36\pm 0.11$, respectively. The slopes of LAEs and LABs are similar within $1\sigma$, while the bright LABs show a $5\sigma$ larger slope than LAEs. This suggests that bright LABs are more clustered than LAEs.

    \item We find that LABs (and bright LABs) have a large galaxy bias of $\sim 5-7$, which suggests a strong field-to-field variance consistent with the large differences of bright LAB number densities in our 8 fields. The large galaxy bias of LABs also suggests that LABs generally reside in more massive dark matter halos (halo masses $M \gtrsim 10^{13}$ M$_{\odot}$) than faint LAEs. 

\end{enumerate}

The Hyper Suprime-Cam (HSC) collaboration includes the astronomical communities of Japan and Taiwan, and Princeton University. The HSC instrumentation and software were developed by the National Astronomical Observatory of Japan (NAOJ), the Kavli Institute for the Physics and Mathematics of the Universe (Kavli IPMU), the University of Tokyo, the High Energy Accelerator Research Organization (KEK), the Academia Sinica Institute for Astronomy and Astrophysics in Taiwan (ASIAA), and Princeton University. Funding was contributed by the FIRST program from Japanese Cabinet Office, the Ministry of Education, Culture, Sports, Science and Technology (MEXT), the Japan Society for the Promotion of Science (JSPS), Japan Science and Technology Agency (JST), the Toray Science Foundation, NAOJ, Kavli IPMU, KEK, ASIAA, and Princeton University. 

This paper makes use of software developed for the Large Synoptic Survey Telescope. We thank the LSST Project for making their code available as free software at  http://dm.lsst.org

The Pan-STARRS1 Surveys (PS1) have been made possible through contributions of the Institute for Astronomy, the University of Hawaii, the Pan-STARRS Project Office, the Max-Planck Society and its participating institutes, the Max Planck Institute for Astronomy, Heidelberg and the Max Planck Institute for Extraterrestrial Physics, Garching, The Johns Hopkins University, Durham University, the University of Edinburgh, Queen’s University Belfast, the Harvard-Smithsonian Center for Astrophysics, the Las Cumbres Observatory Global Telescope Network Incorporated, the National Central University of Taiwan, the Space Telescope Science Institute, the National Aeronautics and Space Administration under Grant No. NNX08AR22G issued through the Planetary Science Division of the NASA Science Mission Directorate, the National Science Foundation under Grant No. AST-1238877, the University of Maryland, and Eotvos Lorand University (ELTE) and the Los Alamos National Laboratory.

Based in part on data collected at the Subaru Telescope and retrieved from the HSC data archive system, which is operated by Subaru Telescope and Astronomy Data Center at National Astronomical Observatory of Japan.

The NB387 filter was supported by KAKENHI (23244022) Grant-in-Aid for Scientific Research (A) through the Japan Society for the Promotion of Science (JSPS).

The authors wish to recognize and acknowledge the very significant cultural role and reverence that the summit of Maunakea has always had within the indigenous Hawaiian community.  We are most fortunate to have the opportunity to conduct observations from this mountain.

We are supported by the National Key R\&D Program of China (grant No. 2018YFA0404503), the National Science Foundation of China (grant No. 12073014), the science research grants from the China Manned Space Project with No. CMS-CSST2021-A05, and the Tsinghua University Initiative Scientific Research Program (No. 20223080023).

\begin{appendix}
\section{Clustering Analysis with K-Function} \label{sec:Kfunc}

\begin{figure}[htb]
\plottwo{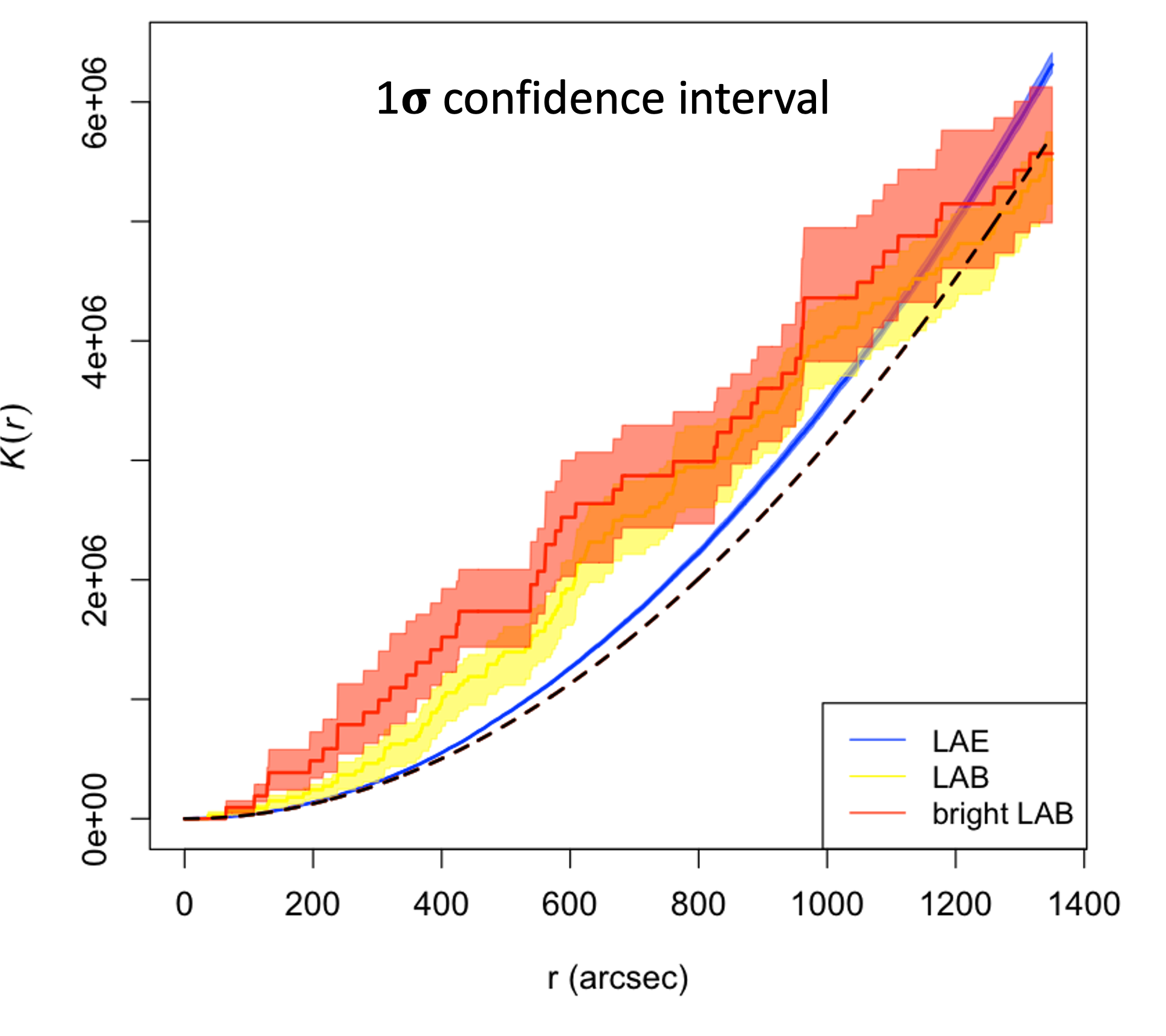}{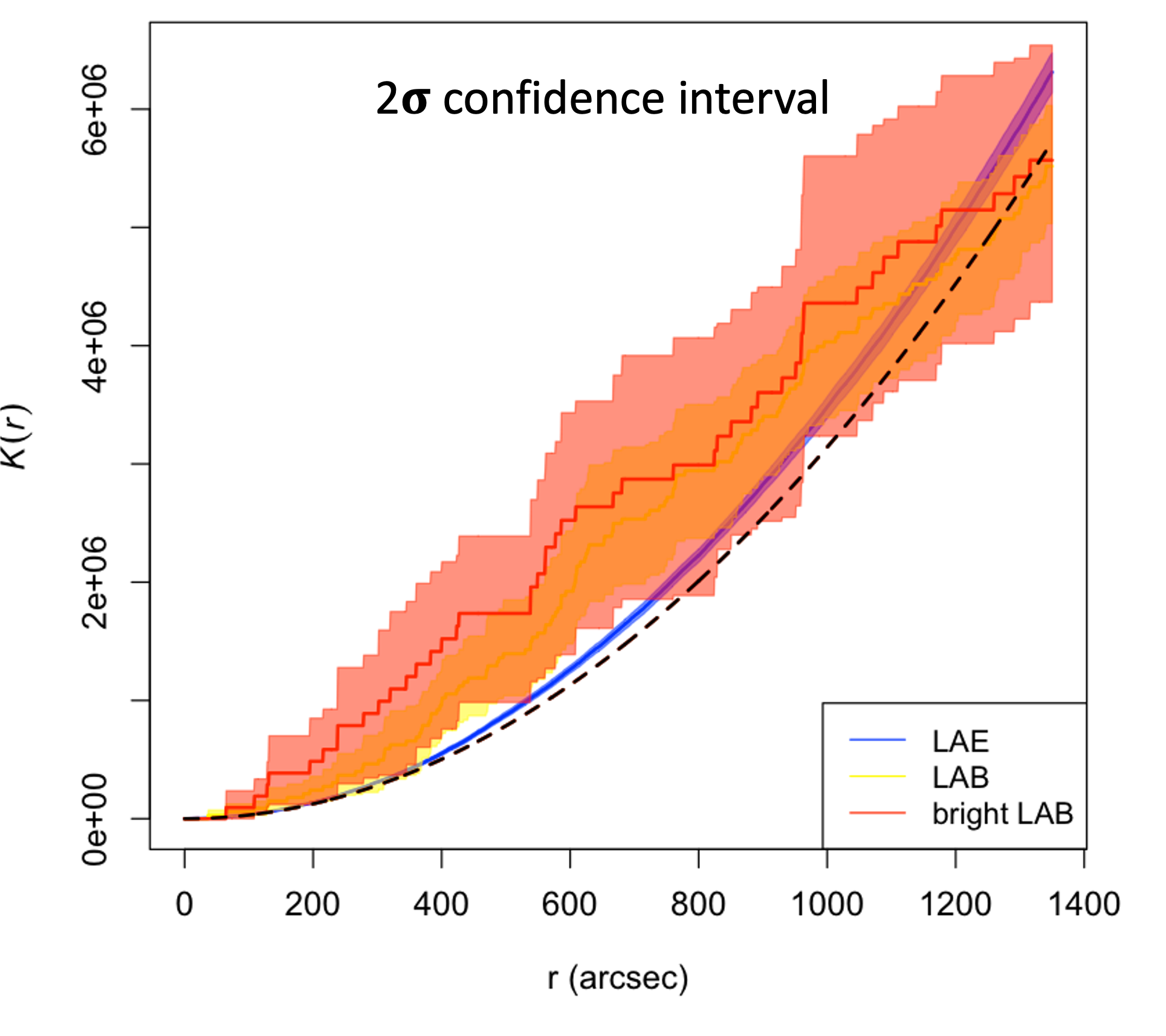}
\caption{K-functions of LAEs (blue), LABs (yellow), and bright LABs (red) in J1349. The black dashed line is the K-function of random Poisson statistics with no clustering. The shaded regions present the $1\sigma$ (68\%; left) and $2\sigma$ (95\%; right) confidence intervals.}
\label{fig:Kfunc}
\end{figure}

In addition to the ACF, we also carry out clustering analysis of LAEs and LABs in J1349 with the K-function (\citealt{Ripley1977}), which measures the standardized cumulative average number of pairs within a distance $r$. We calculate the K-function using the ``Kest" function of the ``spatstat" package (\citealt{Baddeley2015}) in R language. The masked regions are taken into account during the calculation as a ``owin" window defining the observed region. We apply the translation correction for the edge effect (the sampling aperture may extend outside the observed regions such as the masked regions and beyond the field edges). We estimate the errors of K-functions using a bootstrap method (\citealt{Loh2008}; the ``lohboot" function of ``spatstat") with 100 iterations, and measure the $1\sigma$ (68\%) and $2\sigma$ (95\%) confidence intervals. The K-functions of LAEs (with all LABs excluded), LABs, and bright LABs are shown in Figure \ref{fig:Kfunc}.

Apprently, all of the K-functions of LAEs, LABs, and bright LABs are above the random Poisson statistics, which suggest that all of LAEs, LABs, and bright LABs are clustered within $\sim 1000$ arcsec at a $>2\sigma$ significance. Moreover, LABs and bright LABs are more clustered than LAEs within $\sim 1000$ arcsec at a $\sim 2\sigma$ significance. The K-function results are consistent with those from our ACF measurements, suggesting that bright LABs are indeed more clustered than faint LAEs. 

\end{appendix}

\end{document}